\documentclass[article,twocolumn, floatfixnofootinbib]{revtex4}
\usepackage{filecontents}
\usepackage{hyperref}
\usepackage{graphicx} 
\usepackage{subfig}
\usepackage{color}

\newcommand{\ket}[1]{\left| #1 \right\rangle}
\newcommand{\bra}[1]{\left\langle #1 \right|}

\newcommand{\be}{\begin{equation}}
\newcommand{\ee}{\end{equation}}
\newcommand{\bea}{\begin{eqnarray}}
\newcommand{\eea}{\end{eqnarray}}

\usepackage{dcolumn}
\usepackage{bm}
\usepackage{amssymb,amsmath}
\usepackage{color}
\usepackage{float}
\usepackage{tikz}
\usetikzlibrary{arrows}

\definecolor{DarkGreen}{rgb}{0,0.6,0.2}

\begin{document}
\title{Multi-qubit entanglement in bi-directional chiral waveguide QED}
\author{Imran M. Mirza}
\affiliation{Department of Physics, University of Michigan, Ann Arbor, Michigan 48109, USA}
\author{John C. Schotland}
\affiliation{Department of Mathematics and Department of Physics, University of Michigan, Ann Arbor, Michigan 48109, USA}
\begin{abstract}
We study the generation of entanglement induced by a single-photon Gaussian wavepacket in multi-atom bi-directional waveguide QED. In particular, we investigate the effect of increasing the number of atoms on the average pairwise entanglement. We demonstrate by selecting smaller decay rates and in chiral waveguide settings, that both entanglement survival times and maximum generated entanglement can be increased by at least a factor of $\sim 3/2$, independent of the number of atoms. In addition, we analyze the influence of detuning and delays on the robustness of the generated entanglement. There are potential applications of our results in entanglement based multi-qubit quantum networks.
\end{abstract}

\maketitle

\section{Introduction}

Quantum circuits are envisioned to play an indispensable role in the physical implementation of quantum computers~\cite{kimble2008quantum}. In optical quantum computing and in several quantum information processing protocols, controlled light-matter interactions are an essential requirement \cite{ladd2010quantum,northup2014quantum}. Two principal setups have been proposed to achieve such interactions: cavity QED and waveguide QED systems. In cavity QED \cite{mabuchi2002cavity}, matter in the form of qubits interacts with one or a few discrete optical modes confined within an optical resonator. At the same time, atoms can strongly couple with cavity modes thereby producing well-known phenomena such as Rabi oscillations~\cite{horoche1989cavity}. In contrast, in waveguide QED \cite{bermel2006single,zheng2013waveguide}, qubits interact with flying photons which propagate through infinitely many waveguide modes. Such configurations may serve as longer input-output quantum networks. In both types of systems, atom-light interactions can generate qubit-qubit and qubit-photon entanglement, which is a necessary resource for performing many key tasks in quantum information processing and quantum computing.

Waveguide based structures are proving to be excellent platforms for quantum circuits. Some appealing examples in this regard are: plasmonic waveguides \cite{akimov2007generation}, photonic crystals \cite{goban2014atom,chang2013self}, superconducting circuits \cite{rigetti2012superconducting} and optical lattices \cite{beguin2014generation}. In previous waveguide QED studies, two qubit entanglement generation has been analyzed when either an input coherent field or a single photon (produced through an excited qubit) serve as a qubit-qubit entanglement agent \cite{zheng2013persistent, gonzalez2015chiral,gonzalez2011entanglement}. However, an actual quantum network will in general require multiple qubits, wherein flying photons will serve as information carriers.  In this setting,  qubits become entangled at specified nodes in the network.

Motivated by the above considerations, in this paper we study the impact of increasing the number of atoms on single-photon multi-qubit entanglement in bi-directional waveguide QED structures. The theoretical model we consider is  relevant to recent developments in the subject of photonic interactions with a one-dimensional qubit array, mainly in circuit QED and photonic crystal waveguide systems \cite{goban2014atom,fink2009dressed,zhang2014quantum}. We focus specifically on the question of how system parameters can be engineered to control waveguide mediated qubit-qubit entanglement. As opposed to choosing a fixed atom as a single photon source~\cite{gonzalez2015chiral}, here we consider the situation in which a single photon Gaussian wavepacket serves both as an input drive and an entanglement generator. To this end, we derive and then utilize a single-photon bi-directional Fock state master equation. 

The three main approaches used in waveguide QED to study scattering of photons and entanglement are: the real space formalism~\cite{shen2009theory}, the input-output formalism~\cite{fan2010input,caneva2015quantum} and  other master equation approaches \cite{shi2015multiphoton,gonzalez2015chiral}. The main novelty of using the Fock state master equation relies on the fact that it captures both the qubit dynamics and keeps track of the state of the reservoirs at the same time, due to its non-markovian structure. Using this approach, we first study the effect of increasing the number of atoms on the pairwise concurrence. We find that the entanglement survival time markedly decreases. We also find that the maximum concurrence decreases by a factor of $\sim 1/20$ as the number of atoms increases from two to five. However, we demonstrate that small decay rate and chirality can resolve these issues. In addition, we introduce a finite detuning between the peak frequency of the incoming single-photon wavepacket and the atomic transition frequency. We notice that in comparison to the on resonance case, detuning does not affect the overall temporal profile of the entanglement, but the maximum concurrence is reduced. Furthermore, when inter-atomic delays are incorporated, we observe independent of $N$, that smaller delays support an overall larger pairwise concurrence. Moreover, characteristic patterns of death and revival of entanglement appear.

The remainder of this paper is organized as follows. In Sec.~II we introduce the details of the system and its dissipative dynamics. In Sec.~III we report our results.  Finally, in Sec.~IV we close by summarizing our conclusions. In the Appendix, we outline the derivation of the bi-directional master equation that is our main tool in this work.

\section{Theoretical description}
\subsection{Setup}

\begin{figure*}[t]
\includegraphics[width=6.95in,height=1.67in]{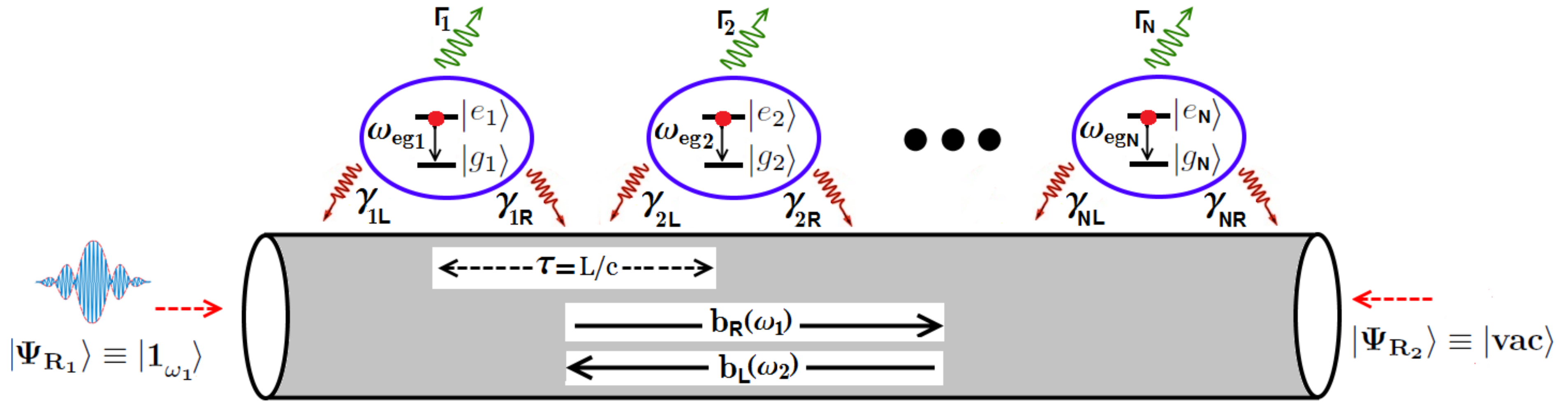}
\captionsetup{
  format=plain,
  margin=1em,
  justification=raggedright,
  singlelinecheck=false
}
 \caption{Illustrating a single-photon wavepacket driving a system of $N$ two-level atoms side coupled to a waveguide. Any two consecutive atoms are separated by the distance $L$ (or time delay $\tau=L/v_{g}$, $v_{g}=c$ being the group velocity of a single photon in the waveguide). Two-mode waveguide continua serve as channels for the wavepacket to propagate through. The atom-waveguide coupling causes atoms to be excited, but also generates qubit-qubit entanglement. The quantity $\Gamma_{i}$ is the emission rate of the $i$th atom into the free space channel; such decays are ignored in the present analysis. Consequently, the coupling fraction parameter~\cite{gonzalez2015chiral} $\beta_{i}=(\gamma_{iL}+\gamma_{iR})/(\gamma_{iL}+\gamma_{iR}+\Gamma_{i})$ has been set equal to unity throughout this paper.}
\label{Fig1}
\end{figure*}
The system under consideration consists of a chain of two-level emitters (atoms, quantum dots, artificial atoms, or Nitrogen vacancy centers in diamond \cite{zoubi2014collective,arcari2014near,fu2008coupling,lalumiere2013input}) side coupled to a dispersionless and lossless waveguide. See Fig.~1.  The frequency of the ground state $\ket{g_{i}}$ and excited state $\ket{e_{i}}$ of the $i$th atom in the chain is denoted by $\omega_{g_{i}}$ and $\omega_{e_{i}}$, for $i=1,\ldots,N$. The process of de-excitation of the $i$th atom is described by the atomic lowering operator $\hat{\sigma}^{-}_{i}=\ket{g_{i}}\bra{e_{i}}$. All atoms are coupled to a common waveguide which has two continua of modes: a left moving continuum and a right moving continuum. Destruction of a single photon in the left (right) moving continuum is described by the annihilation operator $\hat{b}_{L}(\omega_{2})$ $(\hat{b}_{R}(\omega_{2}))$. The waveguide continua are treated as two reservoirs or baths. We will assume that initially the right moving reservoir is in a single-photon pure state $\ket{\Psi_{R_{1}}}$, while the left moving reservoir is in the vacuum state, with $\ket{\Psi_{R_{2}}}=\ket{vac}$. The explicit form of $\ket{\Psi_{R_{1}}}$ is given by
\begin{equation}
\ket{\Psi_{R_{1}}}=\int^{\infty}_{0} g(\omega_{1})\hat{b}^{\dagger}_{R}(\omega_{1})\ket{vac}d\omega_{1} \ ,
\end{equation}
where $g(\omega_{1})$ represents the spectral profile of the single-photon wavepacket. Note that the normalization condition on $\ket{\Psi_{R_{1}}}$ requires that $\int^{\infty}_{0}|g(\omega_{1})|^{2}d\omega_{1}=1$. The non-vanishing commutation relations among operators describing the system are given by
\begin{equation} 
\begin{split}
&[\hat{b}_{R}(\omega_{1}),\hat{b}_{R}(\omega^{'}_{1})]=\delta(\omega_{1}-\omega^{'}_{1})\ ,\\
& [\hat{b}_{L}(\omega_{2}),\hat{b}_{L}(\omega^{'}_{2})]=\delta(\omega_{2}-\omega^{'}_{2})\ , \\
& [\hat{\sigma}^{\dagger}_{i},\hat{\sigma}^{-}_{j}]
=\hat{\sigma}_{zi}\delta_{ij} \ ,
\end{split}
\end{equation} 
where $\hat{\sigma}_{zi}=\ket{e_{i}}\bra{e_{i}}-\ket{g_{i}}\bra{g_{i}}$.
\subsection{Dissipative dynamics and bi-directional Fock state master equation}
The system shown in Fig.~1 is an open quantum system due to the interaction of the atoms with the waveguide continua. However, the dissipative dynamics of the system cannot be described by traditional Born-Markov master equations (Lehmberg type)~\cite{lehmberg1970radiation,pichler2015quantum}. This follows from the fact that once a single photon is absorbed by one of the atoms in the chain, the state of the right moving reservoir changes, which may introduce non-Markovian effects. In view of this observation, we re-derive the single-photon Fock state master equation, which describes the bi-directional coupling between atoms, accounting for decoherence effects. The derivation is outlined in the Appendix A. We thus obtain the following master equation for the evolution of the system density operator $\hat{\rho}_{s}$:
\begin{equation}
\begin{split}
&\frac{d\hat{\rho}_{s}(t)}{dt}= \mathcal{\hat{L}}_{cs}[\hat{\rho}_{s}(t)]+\mathcal{\hat{L}}_{pd}[\hat{\rho}_{s}(t)]+\mathcal{\hat{L}}_{cd}[\hat{\rho}_{s}(t)]\\
&+\sum^{N}_{i=1}\sqrt{2\gamma_{iR}}\left(e^{ik_{0}d_{i}}g(t)[\hat{\rho}_{01}(t),\hat{\sigma}^{\dagger}_{i}]+e^{-ik_{0}d_{i}}g^{\ast}(t)[\hat{\sigma}^{-}_{i},\hat{\rho}_{10}(t)]\right)
\end{split}
\end{equation}
Here for any density operator $\hat{\varrho}(t)$, the action of the aforementioned Liouvillian super operators is given by:
\begin{equation*}
\begin{split}
&\mathcal{\hat{L}}_{cs}[\hat{\varrho}(t)]=-\frac{i}{\hbar}[\hat{H}_{sys},\hat{\varrho}(t)], \hat{H}_{sys}=\hbar\sum^{N}_{i=1}\Delta_{i}\hat{\sigma}^{\dagger}_{i}\hat{\sigma}^{-}_{i} \ , \\
&\mathcal{\hat{L}}_{pd}[\hat{\varrho}(t)]=-\sum^{N}_{i=1}\gamma_{iRL}(\hat{\sigma}^{\dagger}_{i}\hat{\sigma}^{-}_{i}\hat{\varrho}(t)-2\hat{\sigma}^{-}_{i}\hat{\varrho}(t)\hat{\sigma}^{\dagger}_{i}+\hat{\varrho}(t)\hat{\sigma}^{\dagger}_{i}\hat{\sigma}^{-}_{i}) \ , 
\end{split}
\end{equation*}
\begin{equation*}
\begin{split}
&\mathcal{\hat{L}}_{cd}[\hat{\varrho}(t)]=-\sum^{N}_{i\neq j=1}(\sqrt{\gamma_{iR}\gamma_{jR}}\delta_{i>j}+\sqrt{\gamma_{iL}\gamma_{jL}}
\delta_{i<j})\\ \times
&\lbrace(\hat{\sigma}^{\dagger}_{i}\hat{\sigma}^{-}_{j}\hat{\varrho}(t)-\hat{\sigma}^{-}_{i}\hat{\varrho}(t)\hat{\sigma}^{\dagger}_{j})e^{-2\pi i D(i-j)}-(\hat{\sigma}^{-}_{j}\hat{\varrho}(t)\hat{\sigma}^{\dagger}_{i}\\
&-\hat{\varrho}(t)\hat{\sigma}^{\dagger}_{j}\hat{\sigma}^{-}_{i})e^{2\pi i D(i-j)}\rbrace \ ,
\end{split}
\end{equation*}
where $\Delta_{i}=\omega_{egi}-\omega_{p}$ is the detunning between $\omega_{egi}$ and the peak frequency $\omega_{p}$ of the single photon input drive. The parameters $\gamma_{iL}$ and $\gamma_{iR}$ are the spontaneous emission rates of the $i$th atom to decay into the left and right moving waveguide continua, respectively and $\gamma_{iRL}=(\gamma_{iR}+\gamma_{iL})/2$. We also define $k_{0}=\omega_{eg}/v_{g}$ to be the wavenumber of the waveguide emitted photon. Finally, $d_{i}$ specifies the position of the $i$th atom such that $D(i-j)= 2\pi (d_{i}-d_{j})k_{0}$.
The first term on the right hand side of (3) describes the closed system dynamics, the second term represents the pure decay of energy from the atoms into the waveguide continua and the terms multiplied with $\sqrt{\gamma_{iR}\gamma_{jR}}$ and $\sqrt{\gamma_{iL}\gamma_{jL}}$ are the cooperative decay terms (with $j=1,2,3,...,N$).
The operator $\hat{\rho}_{10}$, which appears in (3) is defined as
$$
\hat{\rho}_{10}(t)={\rm Tr}_{R}[\hat{U}(t-t_{0})\hat{\rho}_{s}(t)\ket{vac}\bra{\Psi_{R_{1}}}\hat{\rho}_{R_{2}}(t_{0})\hat{U}^{\dagger}(t-t_{0})] \ .
$$ 
It obeys the equation of motion
\begin{equation}
\begin{split}
&\frac{d\hat{\rho}_{10}(t)}{dt}= \mathcal{\hat{L}}_{cs}[\hat{\rho}_{10}(t)]+\mathcal{\hat{L}}_{pd}[\hat{\rho}_{10}(t)]+\mathcal{\hat{L}}_{cd}[\hat{\rho}_{10}(t)]\\
&+\sum^{N}_{i=1}\sqrt{\gamma_{iR}}e^{-ik_{0}d_{i}}g^{\ast}(t)[\hat{\rho}_{00}(t),\hat{\sigma}^{\dagger}_{i}] \ .
\end{split}
\end{equation}
Here 
$$\hat{\rho}_{00}(t)={\rm Tr}_{R}[\hat{U}(t-t_{0})\hat{\rho}_{s}(t)\ket{vac}\bra{vac}\hat{\rho}_{R_{2}}(t_{0})\hat{U}^{\dagger}(t-t_{0})]$$ 
obeys the no drive (or vacuum) Lehmberg master equation
\begin{equation}
\frac{d\hat{\rho}_{00}(t)}{dt}= \mathcal{\hat{L}}_{cs}[\hat{\rho}_{00}(t)]+\mathcal{\hat{L}}_{pd}[\hat{\rho}_{00}(t)]+\mathcal{\hat{L}}_{cd}[\hat{\rho}_{00}(t)]\ .
\end{equation}

In their study of a continuous mode $\mathcal{N}$-photon wavepacket interacting with a quantum system, Combes et al. have derived a similar master equation for the case $\mathcal{N}=1$~\cite{baragiola2012n}. In their work, they utilized the machinery of quantum stochastic differential equations. We note that the main novelty of our master equation (3) relies on its bi-directional nature, which is more suitable for waveguide QED problems. 
Eqs.~(3), (4) and (5) provide a set of equations needed to obtain a closed form solution for the system density operator $\hat{\rho}_{s}(t)$.
\section{Results and Discussion}
In principle, we can use the bi-directional single-photon Fock state master equation to calculate any observable of interest. In what follows, we will concentrate on how the incident single photon populates the atomic chain, with the concomitant generation of entanglement.  In particular, we will study the evolution of measures of entanglement and the influence of bi-directional waveguide mediated coupling.

\subsection{Influence of the number of atoms on population transfer and pairwise entanglement}
For the remainder of this paper, we assume that the temporal shape of the single-photon wavepacket is a Gaussian function of time of the form
\begin{equation}
g(t)=\frac{1}{\sqrt{2\pi}\Delta t}e^{-(t-\overline{t})^{2}/2(\Delta t)^{2}} \ ,
\end{equation}
where $\overline{t}$ and $\Delta t$ are the mean and width of the Gaussian. Next, as an initial condition, we take all atoms to be in their ground state. That is $\hat{\rho}_{s}(t_{0})=\ket{G}\bra{G}$, $\hat{\rho}_{00}(t_{0})=\ket{G}\bra{G}$ and $\hat{\rho}_{10}(t_{0})=0$ for some initial time $t_0$, where $\ket{G}$ denotes the state in which all atoms occupy their ground state. We also denote by $\ket{E_{1}}$ the state where any one of the atoms is excited. We first calculate the probability $P^{(1)}_{k}(t)=Tr[\hat{\rho}_{s}(t)\ket{E_{1}}\bra{E_{1}}]$ that any one of the atoms in the chain is excited, for $k=1,2,\ldots,5$. We also calculate the corresponding probability that all atoms are in the ground state, denoted $P^{(G)}_{k}(t)=Tr[\hat{\rho}_{s}(t)\ket{G}\bra{G}]$. The main focus of this section will be to investigate how increasing the number of atoms in the chain impacts these probabilities. 
\begin{figure*}
\centering
  \begin{tabular}{@{}cccc@{}}
    \includegraphics[width=3in, height=2.4in]{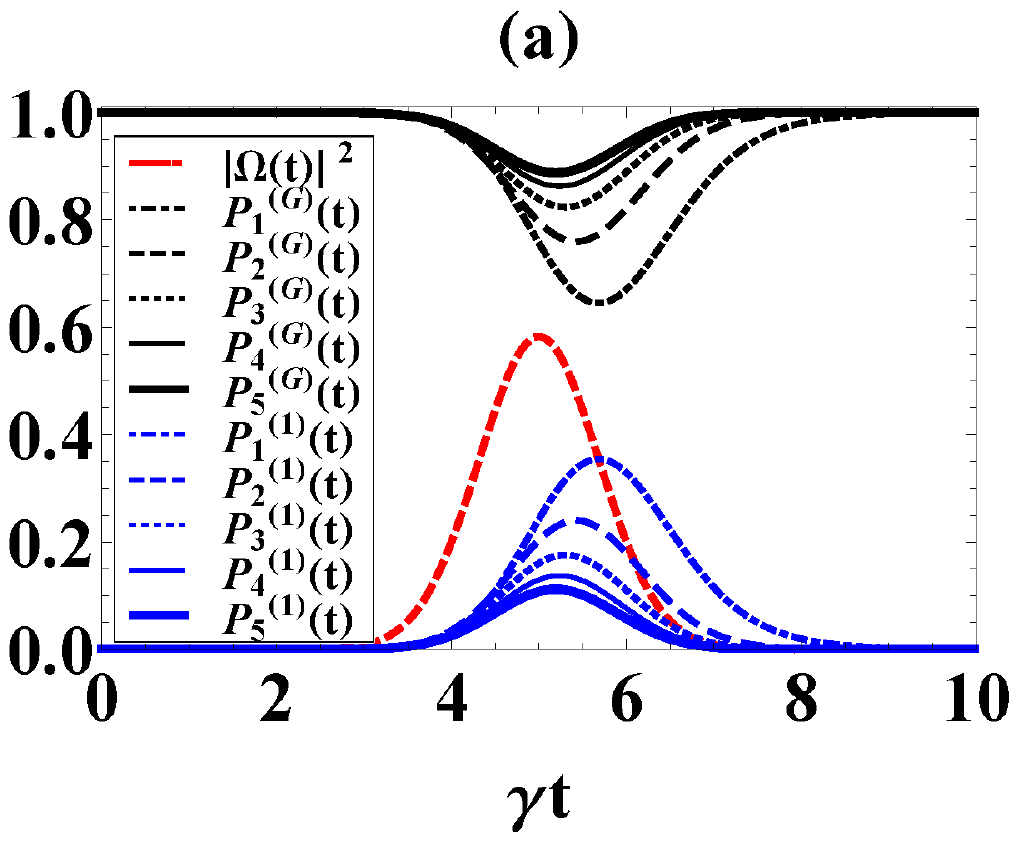}&
    \includegraphics[width=3in, height=2.4in]{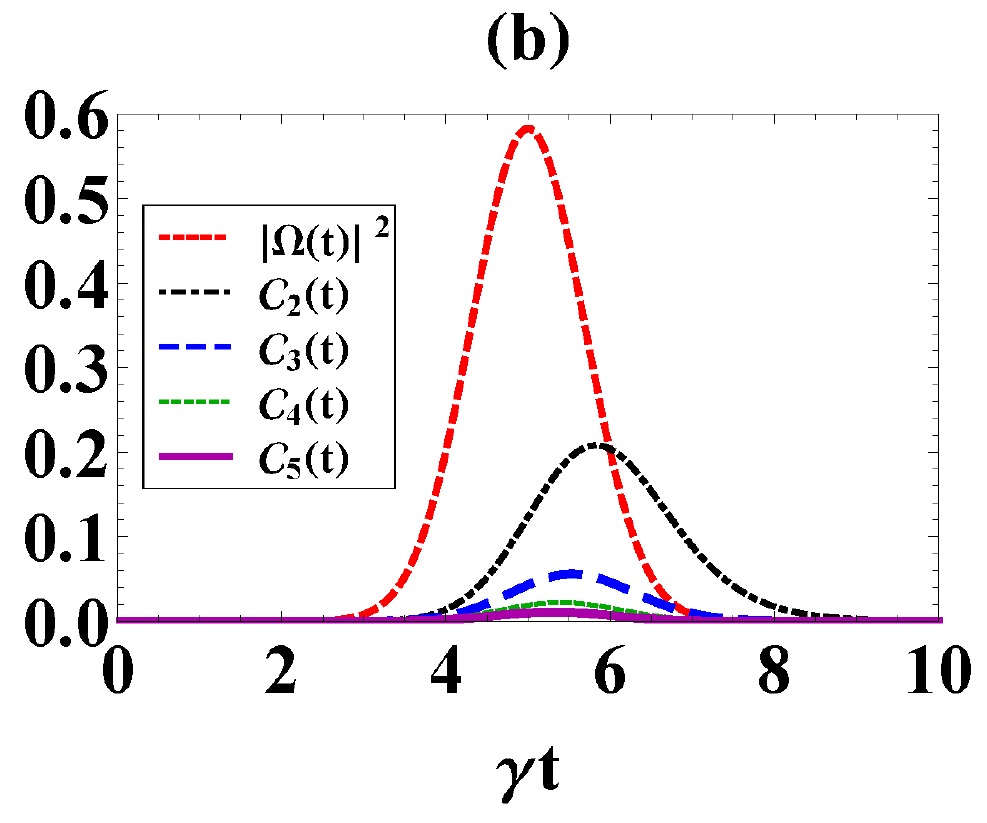}
  \end{tabular}
  \captionsetup{
  format=plain,
  margin=1em,
  justification=raggedright,
  singlelinecheck=false
}
\caption{Time evolution of (a) populations and (b) entanglement as quantified by pairwise concurrence, for a system of up to five atom chain in a waveguide QED setup. A single-photon wavepacket ($\Omega(t)=2\gamma g(t)$) with mean value $5\gamma^{-1}$ and width $1.5\gamma^{-1}$ drives the system strongly ($|\Omega|_{max}>\gamma$) from the right hand side. An on resonance situation is considered with $\omega_{eg}=\omega_{p}$ ($\omega_{eg}$ is the transition frequency of the atoms). For simplicity, all decay rates (pure and cooperative) are assumed to be equal: $\gamma_{iL}=\gamma_{iR}=\gamma$. The single-photon wavepacket parameters are chosen to achieve the maximum value of the single excitation in the system~\cite{wang2011efficient}.}\label{Fig2}
\end{figure*}
Under the above initial conditions and for a single atom in the chain, we obtain a closed form expression for the excitation probability $P^{(1)}_{1}(t)$. To this end, we assume that a single atom is initially unexcited and as an advantageous consequence we observe from Eq.~(5) that the $\hat{\rho}_{00}(t)$ doesn't evolve in time i.e.
\begin{equation}
\hat{\rho}_{00}(t)=e^{\mathcal{L}(t-t_{0})}\hat{\rho}_{00}(t_{0})=\sigma^{-}\sigma^{\dagger} ,
\end{equation}
where, $\hat{\mathcal{L}}[\hat{\varrho}(t)]=\hat{\mathcal{L}}_{cs}[\hat{\varrho}(t)]+\hat{\mathcal{L}}_{pd}[\hat{\varrho}(t)]+\hat{\mathcal{L}}_{cd}[\hat{\varrho}(t)]$. We can then integrate Eq.~(4) to obtain
\begin{equation}
\hat{\rho}_{01}(t)=-\int^{t}_{t_{0}}\Omega^{\ast}(t^{'})e^{(i\omega_{eg}-\gamma)(t-t^{'})}\hat{\sigma}^{-}dt^{'} .
\end{equation}
Inserting the above solution into Eq.~(3) we find the required atom density operator:
\begin{equation}
\hat{\rho}_{s}(t)=\hat{\rho}_{s}(t_{0})+[\hat{\sigma}^{\dagger},\hat{\sigma}^{-}]\int^{t}_{t_{0}}\int^{t^{'}}_{t_{0}}\Omega(t^{'},t^{''})e^{-2\gamma(t-t^{'})}dt^{'}dt^{''} ,
\end{equation}
where $\Omega(t,t^{'})= 2{\rm Re}[\Omega(t)\Omega^{\ast}(t^{'})e^{(i\omega_{eg}-\gamma)(t-t^{'})}]$. To proceed further, we express the temporal profile of the single-photon wavepacket as $\Omega(t)=\mu(t)e^{i\omega_{p}(t)}$, where $\mu(t)$ is assumed to be slowly varying on the time scale of $\gamma^{-1}$. Carrying out the above integral, we obtain
\begin{equation}
\hat{\rho}_{s}(t)\simeq\hat{\rho}_{s}(t_{0})+\Bigg(\frac{2\gamma|g(t)|^{2}}{(\omega_{eg}-\omega_{p})^{2}+\gamma^{2}}\Bigg)[\hat{\sigma}^{\dagger},\hat{\sigma}^{-}] .
\end{equation}
Utilizing this result, the quantity $P^{(1)}_{1}(t)$ can then be obtained.\\

Along with the population dynamics, we will also study the generation and evolution of qubit-qubit entanglement. For the entanglement calculations, we begin with the two-atom chain. For such a bipartite mixed state, the concurrence $\mathcal{C}(\hat{\rho}_{s})$ is a useful measure of entanglement~\cite{wootters1998entanglement}. Following Wootters, we  define the concurrence $\mathcal{C}(t)$ as
\begin{equation}\label{C}
\mathcal{C}(t)=\max\Bigg(0,\sqrt{\lambda_{1}}-\sqrt{\lambda_{2}}-\sqrt{\lambda_{3}}-\sqrt{\lambda_{4}}\Bigg),
\end{equation} 
where $\lambda_{i}$'s are the eigenvalues (in descending order) of the spin-flipped density matrix $\widetilde{\rho}_{s}=\hat{\rho}_{s}(\hat{\sigma}_{y}\otimes\hat{\sigma}_{y})\hat{\rho}^{\ast}_{s}(\hat{\sigma}_{y}\otimes\hat{\sigma}_{y})$, with $\hat{\sigma}_{y}$ being the Pauli spin flip operator. The upper and lower bounds on the concurrence are $1$ and $0$, respectively. We note that $\mathcal{C}=1$ corresponds to a maximally entangled state (for instance, Bell or EPR states), while $\mathcal{C}=0$ corresponds to an unentangled state. For the case of more than two atoms, we will employ the pairwise average concurrence \cite{amico2008entanglement,yonacc2007pairwise,wang2004entanglement,sarovar2010quantum} defined as: $\mathcal{C}(t)=(\sum^{n}_{i=1}\mathcal{C}_{i}(t))/n$, where $n=N/2$ is the total number of pairs of atoms in the chain. We note  that this definition of concurrence has the same properties as each of the individual pair concurrences. \\

We now return to our numerical results. In Fig.~2, we plot the single excitation population dynamics and the temporal profile of the entanglement. We find that a single atom in the chain can be excited with probability $P^{(1)}_{1}$ up to 35\%. This probability remains to one third of its maximum value at the time ($t\sim7\gamma^{-1}$) when the single-photon pulse vanishes.
It takes a further time $t = \gamma^{-1}$ for $P^{(1)}_{1}$ to vanish completely.
This value of $P^{(1)}_{1}$ is less than half of what is reported for a single photon Gaussian input state that is on resonance~\cite{wang2011efficient}. The difference can be attributed to the presence of bi-directional decays in our model. As the number of atoms in the chain increases, we note that the maximum value of population decreases. In particular, for the cases of two, three, four and five atoms in the chain, the maximum population drops down to 24\%, 18\%, 14\% and 11\%, respectively. Moreover, the temporal shape of the excited state populations $P^{(1)}_{k}$ is symmetric about the maximum value induced by the drive.\\
For entanglement calculations, we begin with the case of two qubit concurrence. The spin flip density matrix in this case takes the following form:
\begin{figure*}
\centering
  \begin{tabular}{@{}cccc@{}}
     \includegraphics[width=2.25in, height=2in]{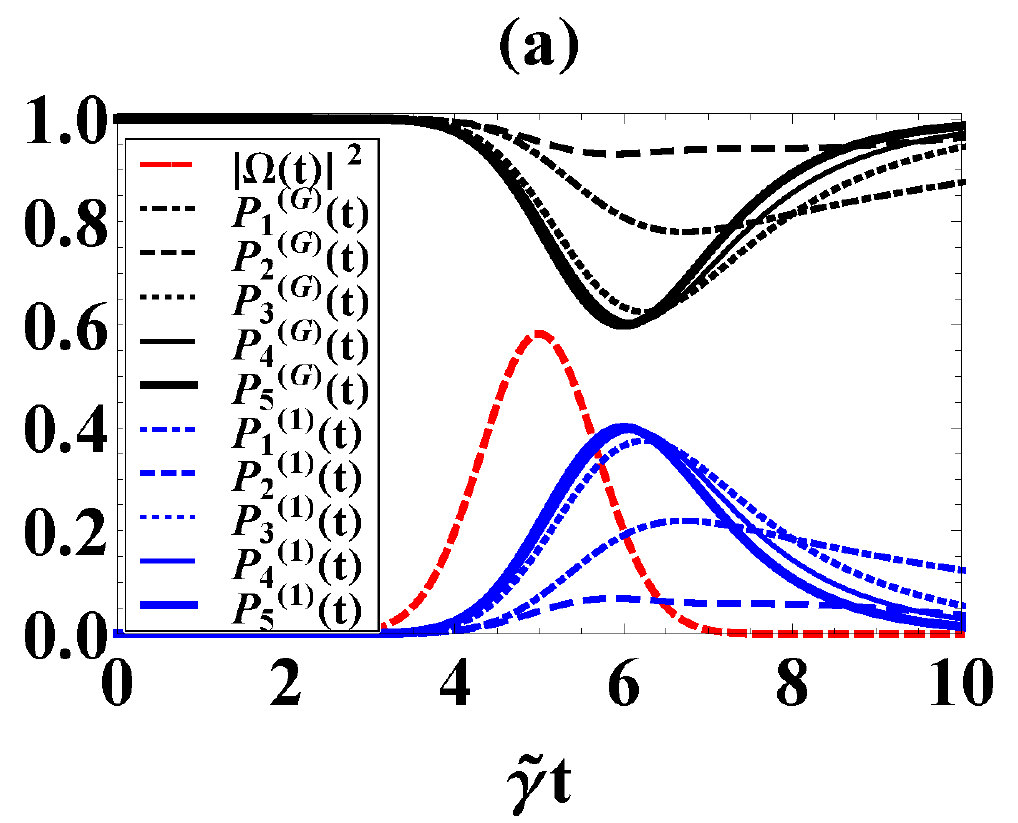}&
    \includegraphics[width=2.25in, height=2.05in]{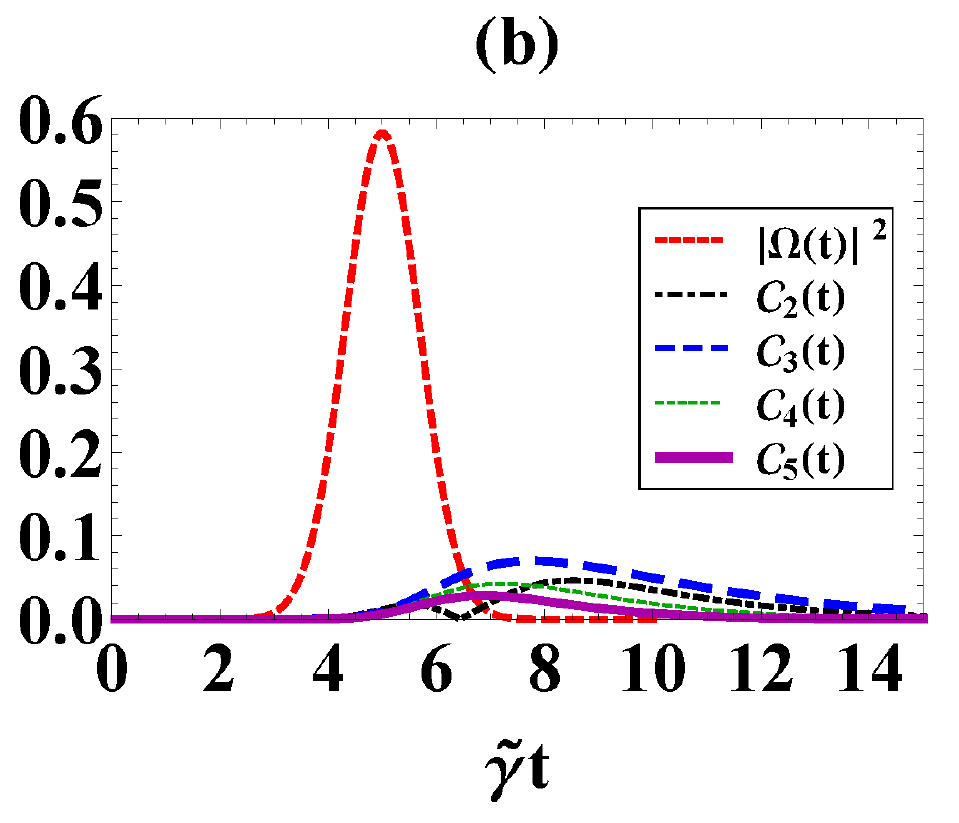}&
    \includegraphics[width=2.3in, height=1.95in]{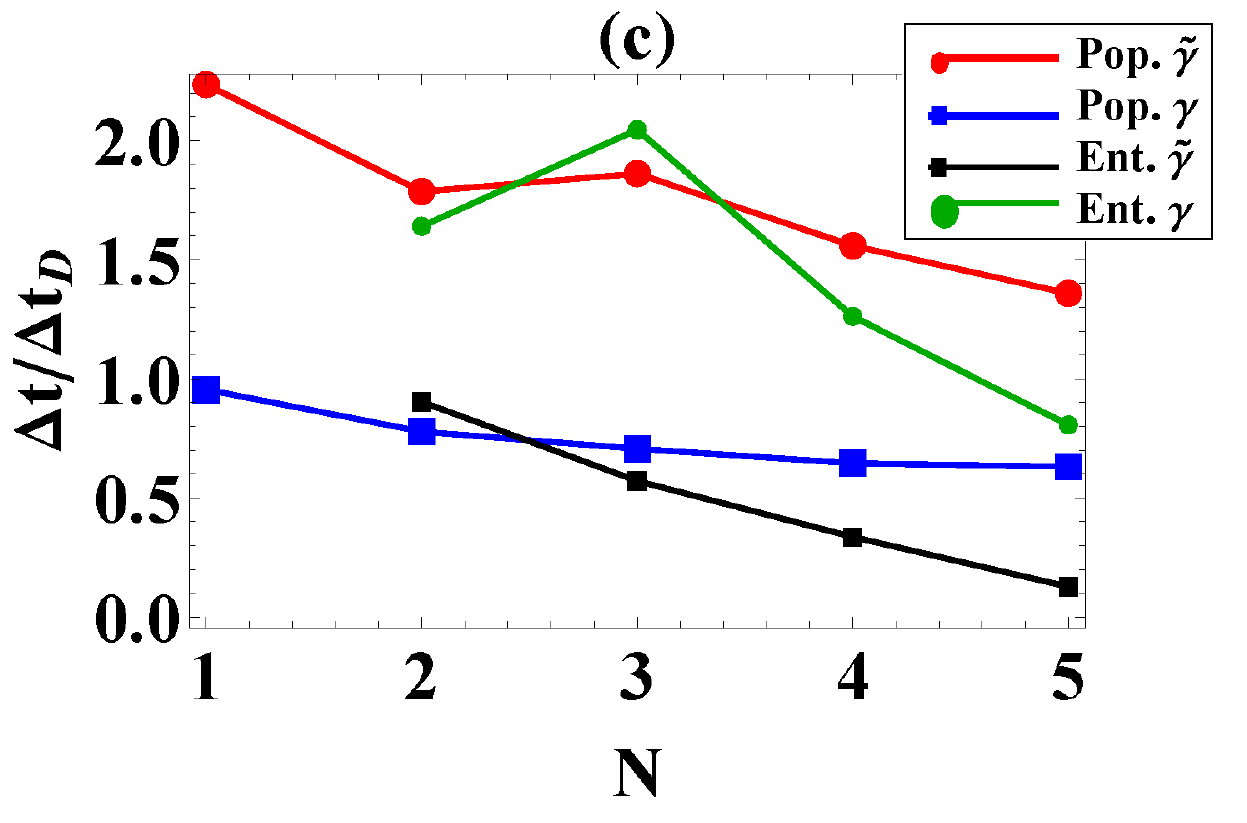}
  \end{tabular}
  \captionsetup{
  format=plain,
  margin=1em,
  justification=raggedright,
  singlelinecheck=false
}
\caption{Effect of small decay rates on the time evolution of (a) populations and (b) pairwise concurrence  for  2, 3, 4 and 5 atoms coupled to a waveguide. All parameters are the same as used in Fig.~2, except that we have chosen smaller cooperative and pure decay rates: $\tilde{\gamma}_{iL}=\tilde{\gamma}_{iR}=\tilde{\gamma}$ while $\tilde{\gamma}=0.1\gamma$. (c) Excited state population and engantlement survival time $\Delta t$ plotted as a function of pulse duration time $\Delta t_{D}$ for both small decay rate $\tilde{\gamma}$ and large decay rate $\gamma$.}\label{Fig3}
\end{figure*}
\begin{equation}
\widetilde{\rho}_{s}(t) = 
 \begin{pmatrix}
  |\rho_{1}|^{2} & 0 & 0 & \rho^{\ast}_{1}\rho_{4} \\
  0 & |\rho_{6}|^{2}+|\rho_{7}|^{2} & |\rho_{6}|^{2}+|\rho_{7}|^{2} & 0 \\
   0 & |\rho_{10}|^{2}+|\rho_{11}|^{2} & |\rho_{10}|^{2}+|\rho_{11}|^{2}& 0  \\
0 & 0 & 0 & 0
 \end{pmatrix} \ ,
\end{equation}
where $\rho_{1}= \bra{g_{1}g_{2}}\hat{\rho}_{s}\ket{g_{1}g_{2}},\rho_{4}= \bra{g_{1}g_{2}}\hat{\rho}_{s}(t)\ket{e_{1}e_{2}}$, $\rho_{6}= \bra{e_{1}g_{2}}\hat{\rho}_{s}(t)\ket{e_{1}g_{2}},\rho_{7}= \bra{e_{1}g_{2}}\hat{\rho}_{s}(t)\ket{g_{1}e_{2}}$, $\rho_{10}= \bra{g_{1}e_{2}}\hat{\rho}_{s}(t)\ket{e_{1}g_{2}}$ and $\rho_{11}= \bra{g_{1}e_{2}}\hat{\rho}_{s}(t)\ket{g_{1}e_{2}}$. Here we employ the notation that the first (second) slot in the ket describes the state of the first (second) atom. We have observed numerically that various entries of $\widetilde{\rho}_{s}(t)$ vanish. We have verified this observation by directly integrating the equation of motion for $\hat{\rho}_{10}(t)$ using the fact that $\hat{\rho}_{00}(t)$ does not evolve in time if both atoms are initially in their ground states. To proceed, we inserted the obtained from of $\hat{\rho}_{10}(t)$ into (3). We found that up to fourth order in $\gamma$, only certain density matrix elements of $\hat{\rho}_{s}(t)$ which appear in (12) survive. Diagonalization of $\widetilde{\rho}_{s}(t)$ then yields the following set of eigenvalues:
\begin{subequations}
\begin{eqnarray*}
\lambda_{1}=0,\lambda_{2}=0,\lambda_{3}=|\rho_{4}|,\\
\lambda_{4}=|\rho_{6}|^{2}+|\rho_{7}|^{2}+|\rho_{10}|^{2}+|\rho_{11}|^{2}=4|\rho_{c}|^{2} \ .
\end{eqnarray*}
\end{subequations}
By numerical integration of (3), (4) and (5), we find that for a system of identical atoms driven by a symmetric Gaussian pulse $\rho_{6}=\rho_{7}=\rho_{10}=\rho_{11}=\rho_{c}$. Inserting these eigenvalues in (7), we obtain a rather compact form of the concurrence: $\mathcal{C}(t)=2\rho_{c}-\rho_{4}$. We have plotted this form of the concurrence in Fig.~2(b).\\

We notice that even in the presence of pure and cooperative decays, an incoming single photon wavepacket generates entanglement between two qubits up to 20.8\%. The entanglement takes $\sim \gamma^{-1}$ time to grow after the initial growth of the input drive. For the present choice of parameters, we find that the atoms remain entangled for a time $5\gamma^{-1}$. As the number of atoms is increased, the pairwise concurrence takes on smaller maximum values. As a result, for the cases of 3, 4 and 5 atoms, the pairwise concurrence attains the values 5.6\%, 2.2\% and 1.1\%, respectively. In addition, the entanglement survives for a corresponding time  of (almost) $4\gamma^{-1}, 3\gamma^{-1}$ and $2.5\gamma^{-1}$.
\subsection{Entanglement storage and small decay rates}
As pointed out above, if we increase the number of atoms in the chain the entanglement is quickly lost. However, for certain quantum information processing protocols, the entanglement survival for prolonged times is one of the key requirements. See Refs.~\cite{clausen2011quantum, saglamyurek2015quantum,ding2015quantum} and the applications mentioned therein. One straightforward way to accomplish this task is to isolate the system from the environment, that is, by setting $\gamma_{iL}=\gamma_{iR}=0$. However, such a choice comes at the price of diminishing qubit-qubit interactions in the system. This includes terms with the pre-factor $\sqrt{\gamma_{iR}\gamma_{jR}},\sqrt{\gamma_{iL}\gamma_{jL}}$ in Eq.~(3), which influences entanglement generation and evolution. Keeping these points in mind, in the present subsection we consider the example of small decay rates. Such rates, for instance, can be obtained in an actual experiment exploiting reservoir engineering techniques \cite{fedortchenko2014finite,schirmer2010stabilizing}.
\begin{figure*}
\centering
  \begin{tabular}{@{}cccc@{}}
    \includegraphics[width=2.25in, height=2in]{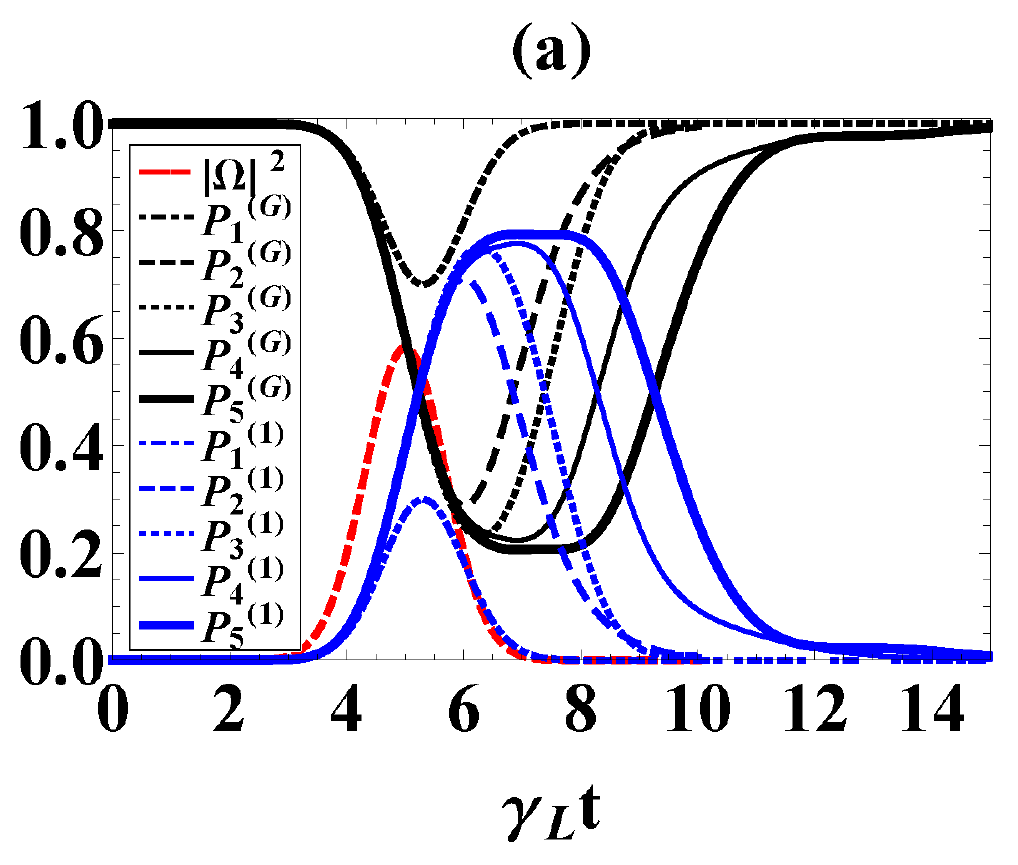}&
    \includegraphics[width=2.25in, height=2.05in]{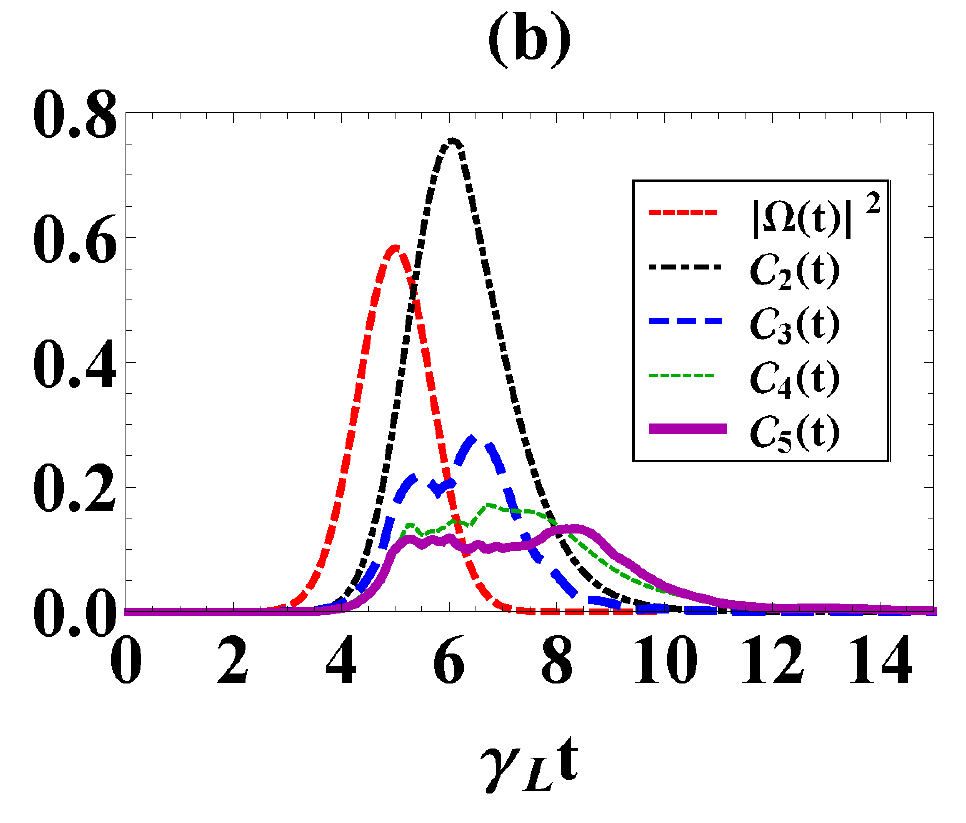}&
    \includegraphics[width=2.25in, height=1.97in]{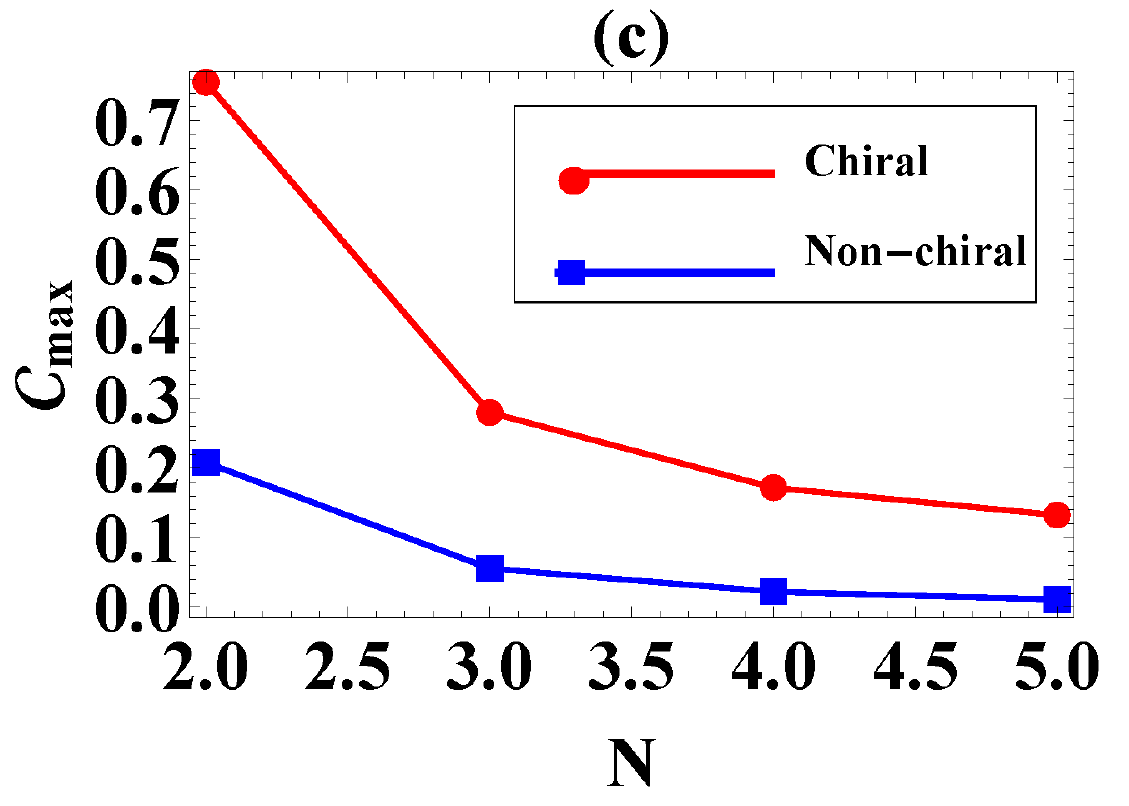}
  \end{tabular}
  \captionsetup{
  format=plain,
  margin=1em,
  justification=raggedright,
  singlelinecheck=false
}
 \caption{Time evolution of (a) populations (b) and pairwise concurrence for a 2, 3, 4 and 5 atoms coupled to a waveguide under chiral condition. We have selected  $\gamma_{1L}=\gamma_{2L}=\gamma_{3L}=\gamma_{4L}=\gamma_{5L}$ $=\gamma_{L}$ (and similarly for all $\gamma_{iR},\forall i=2,3,4,5$) but $\gamma_{iR}/\gamma_{iL}=5$. Other parameters used as used in Fig.~2. (c) Comparison of maximum entanglement $\mathcal{C}_{max}$ achieved under the present chiral setting as compared to non-chiral case, both plotted as a function of number of atoms ($N$) in the atomic chain.}\label{Fig4}
\end{figure*}

In Figs.~3(a) and (b), we consider small decay rates, which results in a longer survival of both the single-excitation populations as well as the pairwise concurrence among atoms. The highest values achieved by the population almost remain the same as found in Fig.~2, but the entanglement tends to achieve smaller maximum for a bipartite system. Moreover, the case of two atoms also shows the phenomenon of entanglement death and revival \cite{mazzola2009sudden,xu2010experimental}. However, when three, four or five atoms are included in the chain, the highest values of the entanglement for both small decay rates ($\tilde{\gamma}=0.1\gamma$) and large decay rates ($\gamma$) almost matches. 

The survival times (in terms of the pulse duration time $\Delta t_{D}=4.85\gamma^{-1}$) are plotted in Fig.~3(c) as a function of the number of atoms in the chain, for the cases of both small and large decay. We find that as soon as we choose $\tilde{\gamma}$ as the decay rate, both populations ($\Delta t_{p}$) and entanglement ($\Delta t_{\mathcal{C}}$) survival times increase by a factor of $2$ compared to a decay rate of $\gamma$. Another interesting feature in Fig.~3(c) is a jump in $\Delta t_{p}$ and $\Delta t_{\mathcal{C}}$  as we move from 2 to 3 atom chains. This behavior seems to  indicate that there is an optimal point between the extremes of pure decay and qubit-qubit coupling. 
In the case of 3 atoms in the chain, qubit-qubit coupling starts to dominate for smaller decay rates. In the 4 and 5 atoms cases, decay mechanisms begin to overwhelm inter-atomic couplings, which results in the same behavior followed by the curves with larger decay rates.
\subsection{Chiral atom-waveguide couplings}
Recently~\cite{petersen2014chiral} it has been shown that the spin-orbit interaction of light leads to symmetry breaking in the atomic-emission direction in waveguide QED. The resulting waveguide system is known as a quantum chiral network. Along with theoretical efforts~\cite{pichler2015quantum,ramos2014quantum}, many experiments have been performed to study chirality effects in various systems~\cite{petersen2014chiral, le2014anisotropy, sollner2015deterministic}. In particular, in photonic crystals~\cite{sollner2015deterministic}  90\% directionalities and 98\% atom-waveguide coupling strengths have been achieved.
In view of these developments, in this subsection we suppose that the emission from all atoms is preferential in one direction. To this end, we take $\gamma_{iR}=5\gamma_{iL}$, $i=1,\dots,5$, while utilizing the fact that the single-photon drive is also launched towards the right in the waveguide which enhances the interaction of the first atom with its partners towards right. Note that these chiral decay rates values and $\beta\sim 1$ lies within the experimental reported values \cite{sollner2015deterministic}. 

In Fig.~4(a) we plot the population dynamics. We observe that the effect of chiralty is marked as compared to the non-chiral setting shown in Fig.~2. Chirality supports better single excitation transfer to the system. It also supports longer population trapping as the number of atoms is increased. This can be quantitatively understood by noticing that compared to single atom case, in the case of two, three, four and five atoms, the maximum population attained by the system becomes 24.7\%, 25.7\%, 25.9\% and  26.5\%, respectively. Besides the longer overall  survival of a single photon in the system as a function of the number of atoms, a plateau emerges around the maximum value of the single excitation population, becoming more pronounced as the number of atoms are increased.

As shown in Fig.~4(b), chirality also enhances entanglement. For the two qubit case, this enhancement is more than three times greater than the corresponding non-chiral case (see Fig.~4(c)). For the multi-qubit cases, the maximum pairwise concurrence remains 3/2 times as large as in the non-chiral case. In such multipartite situations, we also note the appearance of an oscillatory pattern in $\mathcal{C}$. Such a pattern, which eventually turns into an entanglement plateau, exhibits the fact that with a larger number of atoms in the system, a single photon transfers back and forth among qubits with unequal probability, such that the overall pairwise entanglement survives for an extended period of time.
\subsection{Detuning and Delays}
 \begin{figure*}[t]
\centering
  \begin{tabular}{@{}cccc@{}}
    \includegraphics[width=3in, height=2.53in]{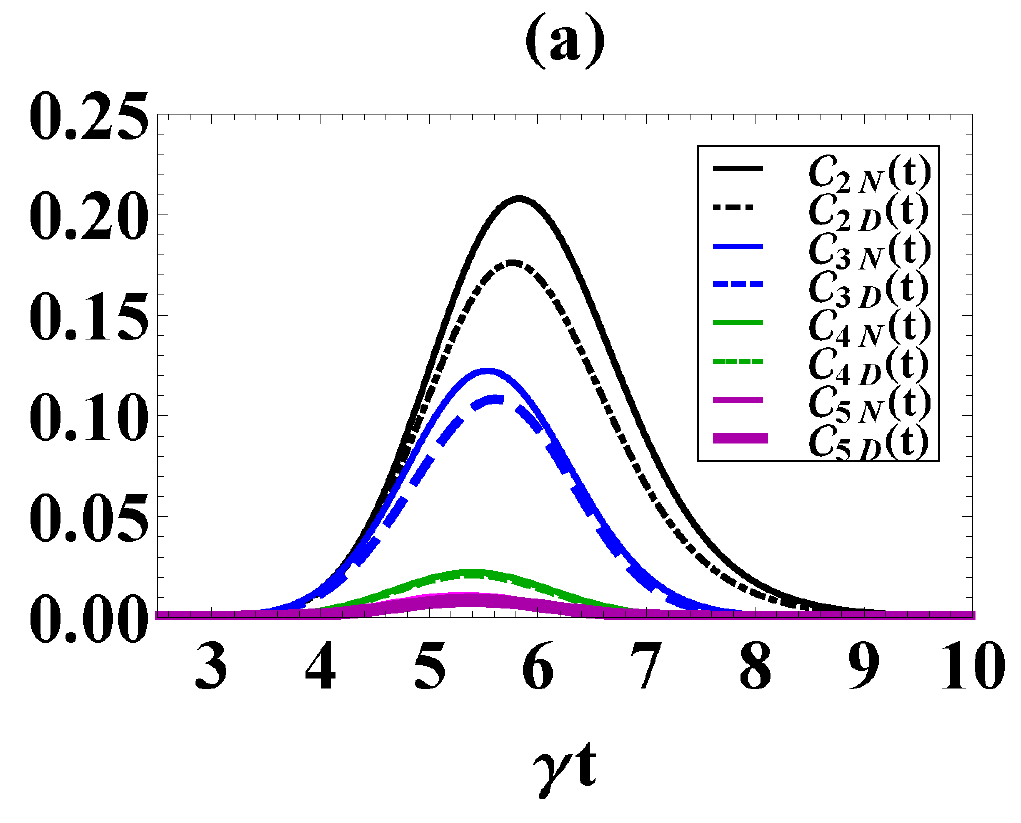}&
    \hspace{5mm}\includegraphics[width=3in, height=2.4in]{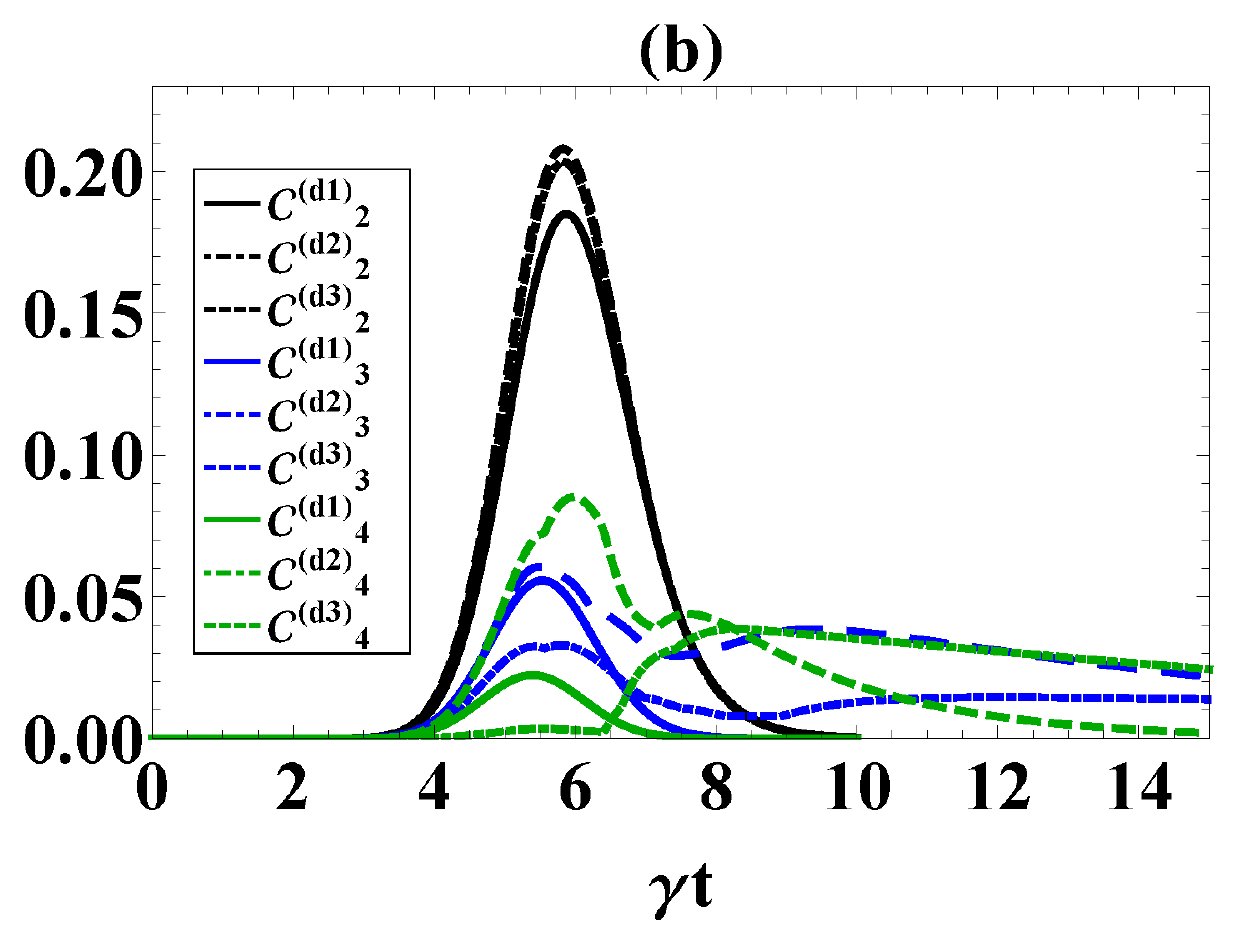}
  \end{tabular}
  \captionsetup{
  format=plain,
  margin=1em,
  justification=raggedright,
  singlelinecheck=false
}
  \caption{(a) Finite detunning and entanglement evolution. All atoms in the chain are assumed to have the same resonant frequency $\omega_{eg}$, which is $0.5\gamma$ detuned from $\omega_{p}$. Here $\mathcal{C}_{kN}$ and $\mathcal{C}_{kD}$ are the concurrence for no-detuning and finite detuning cases, respectively for $k=2,3,4,5.$ (b) Time delays between the atoms and entanglement evolution. Three different phases are plotted namely $d_{1}=L, d_{2}=L/8$ and $d_{3}=L/16$ (while $D\lambda_{0}$ used in Eq.~3 equals $d_{1},d_{2}$ and $d_{3})$. The remaining parameters (for both part (a) and (b)) are the same as in Fig.~2.}\label{Fig5}
\end{figure*} 
We now suppose that the incoming single-photon is detuned from the atomic resonances, that is $\omega_{p}$ is slightly mismatched from $\omega_{eg}$. In Fig. 5(a) we present our numerical results. We notice that the detuning does not affect the overall profile of the pairwise concurrence. Moreover, as the number of atoms increases, the difference between the $\mathcal{C}_{max}$values attained by detunned and on resonance cases becomes smaller.  \\
Next, we consider the effect of delays on entanglement. The delays we consider are introduced through the phases appearing in the atom-waveguide interaction Hamiltonian (see Eq.~A1). Three different phases (inter-atomic separations) are considered, namely $L/\lambda_{0},L/16\lambda_{0}$ and $L/8\lambda_{0}$.
Even though the structure of the master equation at hand is non-Markovian, we impose the Markovian regime requirement on delays i.e. $\gamma D \leq v_{g}$ \cite{fang2015waveguide,tufarelli2014non}. 
In Fig. 5(b) we have plotted the quantity $\mathcal{C}(t)$ including the effects of delays.
In the two-atom case, we observe that as we decrease the separation from $L$ to $L/8$ (dashed black curve) and finally to $L/16$ (dotted black curve in Fig. 5(b)), the entanglement exhibits a slight enhancement.
In the three atom case, we point out that as the separation is reduced, the entanglement shows two
regions of growth and decay. For $d = L/8, L/16$, the entanglement shows a partial decay
after an initial growth, while later in time the entanglement decays slowly. Similarly, in the 4
atom case, the smallest separation produces the largest maximum entanglement (for 3 atoms
$\sim 0.06$ while for four atoms it becomes $\sim 0.08$). For the four atom example, the entanglement
is more than three times the maximum entanglement gained for the case of the largest separation ($\sim 0.024$). This behavior suggests that by decreasing the distance between the atoms, the width of the photonic wave packet emitted by the first atom becomes larger than the qubit-qubit separation. As a result, before the decay of the first qubit, the population reaches the second qubit, and from the second qubit this process extends to the third qubit and so on. Hence, the overall concurrence becomes more pronounced with increasing number of qubits. Finally, we remark that the revival profile of the pairwise concurrence that is observed for smaller separations provides a means to probe the temporal pattern of entanglement by varying the atomic separation.

\section{Conclusions}
In this paper, we have studied the manner in which a single-photon wavepacket with a Gaussian spectral profile can distribute its population and stimulate entanglement among atoms in lossless waveguide QED. By applying a bi-directional single-photon Fock state master equation, we report several findings. First, as the number of atoms increases, both the single-excitation population as well as the average pairwise concurrence are considerably reduced. Second, the problem of short entanglement survival time is somewhat mitigated by the utilization of small decay rates. Third, we have found that the introduction of chirality can increase the entanglement and population by more than a factor of 3/2 compared to the non-chiral case. Fourth, nonzero detuning has only a modest effect on entanglement. Inclusion of smaller delays leads to higher maximum entanglement. Finally, entanglement death and revival patterns appear which allow some control of the overall temporal profile of the entanglement. Such control is important for practical implementation of the proposed model.

\section*{Acknowledgments}
This work was supported by the NSF Grants DMR-1120923, DMS-1115574 and DMS-1108969.
\setcounter{equation}{0}
\makeatletter
\section*{Appendix A: Derivation of bi-directional single-photon Fock state Master equation}
\renewcommand{\theequation}{A\arabic{equation}}
We decompose the $N$-system chain into $N$ subsystems. The dissipative dynamics of the first subsystem can be described in the Heisenberg picture through the following quantum Langevin equation \cite{weiss1999quantum,gardiner2004quantum}:
\begin{equation}
\begin{split}
&\frac{d\hat{X}_{1}(t)}{dt}=\frac{-i}{\hbar}[\hat{X}_{1}(t),\hat{H}_{sys1}]-\\
&[\hat{X}_{1}(t),\hat{c}^{\dagger}_{1}(t)]\Bigg(\sqrt{\gamma_{1R}}e^{ik_{0}d_{1}}\hat{b}^{(1R)}_{in}(t)
+\sqrt{\gamma_{1L}}e^{-ik_{0}d_{1}}\hat{b}^{(1L)}_{in}(t)\\
&+(\frac{\gamma_{1R}+\gamma_{1L}}{2})\hat{c}_{1}  \Bigg)+{\rm h.c.} \ ,
\end{split}
\end{equation}
where $\hat{X}_{1}(t)$ and $\hat{c}_{1}(t)$ are arbitrary Heisenberg picture operators belonging to system-1 and ${\rm h.c.}$ stands for the hermitian conjugate of the terms whose prefactor is the commutator $[\hat{X}_{1}(t),\hat{c}^{\dagger}_{1}]$. In writing this equation, we have identified two ``input" operators:
\begin{subequations}
\begin{eqnarray}
\hat{b}_{in}^{(1R)}(t)=\frac{1}{\sqrt{2\pi}}\int^{\infty}_{-\infty}\hat{b}_{R}(\omega_{1},t_{0})e^{-i\omega_{1}(t-t_{0})}d\omega_{1} \ , \\
\hat{b}_{in}^{(1L)}(t)=\frac{1}{\sqrt{2\pi}}\int^{\infty}_{-\infty}\hat{b}_{L}(\omega_{2},t_{0})e^{-i\omega_{2}(t-t_{0})}d\omega_{2} \ ,
\end{eqnarray}
\end{subequations}
where $t_{0}$ represents an initial time, which can be set equal to zero without loss of generality. The input operators obey the causality condition as indicated by the commutation relation: $[\hat{b}^{(1j)}_{in}(t),\hat{b}^{\dagger(1j)}_{in}(t^{'})]=\delta(t-t^{'}), \ j=R,L$. Following along the same lines, one can express the dissipative dynamics of each individual subsystem through a similar Langevin equation.

To combine the independent Langevin equations for each atom, we note that for each of the input operators appearing in Eq.~A2, there exist two output operators. For subsystem 1 these input and output operators are linked through the input-output relations \cite{gardiner2004quantum}:
\begin{subequations}
\begin{eqnarray}
\hat{b}^{(1R)}_{out}(t)=\hat{b}^{(1R)}_{in}(t)+\sqrt{\gamma_{1R}}e^{-ik_{0}d_{1}}\hat{c}_{1}(t)\ , \\
\hat{b}^{(1L)}_{out}(t)=\hat{b}^{(1L)}_{in}(t)+\sqrt{\gamma_{1L}}e^{ik_{0}d_{1}}\hat{c}_{1}(t) \ ,
\end{eqnarray}
\end{subequations}
where $t_{1}$ is a future time. We define the output operators as
\begin{subequations}
\begin{eqnarray}
\hat{b}_{out}^{(1R)}(t)=\frac{1}{\sqrt{2\pi}}\int^{\infty}_{-\infty}\hat{b}_{R}(\omega_{1},t_{1})e^{-i\omega_{1}(t-t_{1})}d\omega_{1} \ , \\
\hat{b}_{out}^{(1L)}(t)=\frac{1}{\sqrt{2\pi}}\int^{\infty}_{-\infty}\hat{b}_{L}(\omega_{2},t_{1})e^{-i\omega_{2}(t-t_{1})}d\omega_{2} \ .
\end{eqnarray}
\end{subequations}
Next, we note that the output from one subsystem feeds into the nearest subsystems as a time-delayed input. For instance, for just two subsystem example we have
\begin{subequations}
\begin{eqnarray}
\hat{b}_{in}^{(2R)}(t)=\hat{b}_{out}^{(1R)}(t-\tau)=\hat{b}_{in}^{(1R)}(t-\tau)+\sqrt{\gamma_{1R}}e^{-ik_{0}d_{1}}\hat{c}_{1}(t-\tau) \ , \nonumber\\
\hat{b}_{in}^{(1L)}(t)=\hat{b}_{out}^{(2L)}(t-\tau)=\hat{b}_{in}^{(2L)}(t-\tau)+\sqrt{\gamma_{1R}}e^{ik_{0}d_{1}}\hat{c}_{2}(t-\tau)\nonumber \ .
\end{eqnarray}
\end{subequations} 
If we neglect the time-delays, assuming that each subsystem evolves on a time scale much slower than the time a photon takes to travel between the subsystems:  $\omega_{egi},\gamma_{il}\ll 1/\tau={L}/{c}, \ l=R,L$, we arrive at the following bi-directional combined Langevin equation for an arbitrary operator $\hat{X}(t)$:
\begin{equation}
\begin{split}
&\frac{d\hat{X}(t)}{dt}=\frac{-i}{\hbar}[\hat{X},\hat{H}_{sys}]\\
&-\sum^{N}_{i=1}\Bigg\lbrace[\hat{X},\hat{c}^{\dagger}_{i}]\Bigg(\sqrt{\gamma_{iR}}e^{ik_{0}d_{i}}\hat{b}^{(iR)}_{in}+\sqrt{\gamma_{iL}}e^{-ik_{0}d_{i}}\hat{b}^{(iL)}_{in}\\
&+(\frac{\gamma_{iR}+\gamma_{iL}}{2})\hat{c}_{i}
+\sum^{N}_{j\neq i=1}e^{ik_{0}(d_{i}-d_{j})}(\sqrt{\gamma_{iR}\gamma_{jR}}\delta_{i>j}\hat{c}_{j}\\
&+\sqrt{\gamma_{iL}\gamma_{jL}}\delta_{i<j}\hat{c}_{j})\Bigg)+{\rm h.c.}\Bigg\rbrace \ .
\end{split}
\end{equation}
Here bidirectionality is  manifested by terms with prefactors $\sqrt{\gamma_{il}\gamma_{jl}}, \ l=R,L$ and $\delta_{i\lessgtr j}=1$, only when $i\lessgtr j$. Next, we transform to the Schr\"odinger picture using the identity: 
\be
{\rm Tr}_{S\oplus R}\left[\frac{d\hat{X}(t)}{dt}\hat{\rho}_{s}(t_{0})\right]={\rm Tr}_{S}\left[\hat{X}(t_{0})\frac{d\hat{\rho}_{s}(t)}{dt}\right] \ ,
\ee
where $\hat{\rho}_{s}(t)$ is the system reduced density matrix. Therefore, we obtain
\begin{equation}
\begin{split}
&\frac{d\hat{\rho}_{s}(t)}{dt} = \mathcal{\hat{L}}_{cs}[\hat{\rho}_{s}(t)]+\mathcal{\hat{L}}_{pd}[\hat{\rho}_{s}(t)]+\mathcal{\hat{L}}_{cd}[\hat{\rho}_{s}(t)]\\
&-{\rm Tr}_{S\oplus R}\Bigg[\sum^{N}_{i=1}\Bigg(\sqrt{\gamma_{iR}}(e^{ik_{0}d_{i}}[\hat{X}(t),\hat{c}^{\dagger}_{i}(t)]\hat{b}^{(1R)}_{in}(t)\hat{\rho}(t_{0})\\
&-e^{-ik_{0}d_{i}}\hat{b}^{\dagger(1R)(t)}_{in}[\hat{X}(t),\hat{c}_{i}(t)]\hat{\rho}(t_{0}))-\sqrt{\gamma_{iL}}(e^{-ik_{0}d_{i}}[\hat{X}(t),\hat{c}^{\dagger}_{i}(t)]\\
&\times\hat{b}^{(NL)}_{in}(t)\hat{\rho}(t_{0}) -e^{ik_{0}d_{i}}\hat{b}^{\dagger(NL)(t)}_{in}[\hat{X}(t),\hat{c}_{i}(t)]\hat{\rho}(t_{0}))\Bigg)\Bigg] \ ,
\end{split}
\end{equation}
where $D=L/\lambda_{0}$. We now focus our attention on the input operator terms. We note that a considerable simplification arises from the fact that the state of the left moving continuum is initially the vacuum. As a result, all terms involving the $\hat{b}^{(NL)}_{in}(t)$ operator  must vanish:
\begin{equation*}
\begin{split}
&{\rm Tr}_{S\oplus R}\Bigg[[\hat{X}(t),\hat{c}^{\dagger}_{i}(t)]\hat{b}^{(iL)}_{in}(t)\hat{\rho}(t_{0})\Bigg]=\\
&{\rm Tr}_{S\oplus R}\Bigg[[\hat{X}(t),\hat{c}^{\dagger}_{i}(t)]\hat{\rho}_{s}(t_{0})\otimes\hat{\rho}_{R1}(t_{0})\otimes\hat{b}^{(iL)}_{in}(t)\ket{vac}\bra{vac}\Bigg]
\\
& =0 \ ,
\end{split}
\end{equation*}
where we have assumed that the initial state of the global system is factorizable into system and bath initial states. Note that the right moving continuum input terms does not vanish due to the presence of a single photon in the initial state of the reservoir.

For the single-photon wavepacket in (1), we find that  $\hat{b}^{(iR)}_{in}(t)\ket{\Psi_{R_{1}}}=g(t)\ket{vac}$ and hence
\begin{equation*}
\begin{split}
&{\rm Tr}_{S\oplus R}\Bigg[[\hat{X}(t),\hat{c}^{\dagger}_{i}]\hat{b}^{(iR)}_{in}(t)\hat{\rho}(t_{0})\Bigg]=\\
&{\rm Tr}_{S\oplus R}\Bigg[[\hat{X}(t),\hat{c}^{\dagger}_{i}(t)]\hat{\rho}_{s}(t_{0})\otimes\hat{b}^{(iR)}_{in}(t)\ket{\Psi_{R_{1}}}\bra{\Psi_{R_{1}}}\otimes\hat{\rho}_{R2}(t_{0})\Bigg] \\
&=g(t){\rm Tr}_{S}[\hat{X}(t_{0})[\hat{c}^{\dagger}_{i},\hat{\rho}_{01}(t)]] \ ,
\end{split}
\end{equation*}
with $g(t)$ being the temporal shape of the single-photon wavepacket. The density matrix element $\hat{\rho}_{01}(t)$ is a novel and  a non-physical density operator; it follows that $\hat{\rho}^{\dagger}_{01}(t)=\hat{\rho}_{10}(t)$. The form of $\hat{\rho}_{10}(t)$ has already been mentioned in Sec.~II(B). Putting everything together, we obtain the required bi-directional single photon Fock state master equation
\begin{equation}
\begin{split}
&\frac{d\hat{\rho}_{s}(t)}{dt}= \mathcal{\hat{L}}_{cs}[\hat{\rho}_{s}(t)]+\mathcal{\hat{L}}_{pd}[\hat{\rho}_{s}(t)]+\mathcal{\hat{L}}_{cd}[\hat{\rho}_{s}(t)]+\\
&\sum^{N}_{i=1}\sqrt{2\gamma_{iR}}(e^{ik_{0}d_{i}}g(t)[\hat{\rho}_{01}(t),\hat{\sigma}^{\dagger}_{i}]+e^{-ik_{0}d_{i}}g^{\ast}(t)[\hat{\sigma}^{-}_{i},\hat{\rho}_{10}(t)])
\ .
\end{split}
\end{equation}
In order to obtain the equation of motion obeyed by ${\rho}_{10}$, we use the identity mentioned in (A7) to obtain
\be
{\rm Tr}_{S\oplus R}\left[\frac{d\hat{X}(t)}{dt}\hat{\rho}_{10}(t_{0})\right]={\rm Tr}_{S}\left[\hat{X}(t_{0})\frac{d\hat{\rho}_{10}(t)}{dt}\right] \ .
\ee 
Consequently, we find that
\begin{equation}
\begin{split}
&\frac{d\hat{\rho}_{10}(t)}{dt}= \mathcal{\hat{L}}_{cs}[\hat{\rho}_{10}(t)]+\mathcal{\hat{L}}_{pd}[\hat{\rho}_{10}(t)]+\mathcal{\hat{L}}_{cd}[\hat{\rho}_{10}(t)]\\
&+\sum^{N}_{i=1}\sqrt{\gamma_{iR}}e^{-ik_{0}d_{i}}g^{\ast}(t)[\hat{\rho}_{00}(t),\hat{\sigma}^{\dagger}_{i}] \ .
\end{split}
\end{equation}
Likewise, we see that $\hat{\rho}_{00}(t)$ obeys
\begin{equation}
\frac{d\hat{\rho}_{00}(t)}{dt}= \mathcal{\hat{L}}_{cs}[\hat{\rho}_{00}(t)]+\mathcal{\hat{L}}_{pd}[\hat{\rho}_{00}(t)]+\mathcal{\hat{L}}_{cd}[\hat{\rho}_{00}(t)] \ .
\end{equation}

\bibliographystyle{ieeetr}
\bibliography{Article}
\end{document}